\documentstyle[dina4my,12pt,epsfig]{article}
\newcommand{\PO}{\em I\! \! P }
\newcommand{\pom}{I\!\!P}
\newcommand{\xpom}{x_{\pom}}
\newcommand{\JP}{J/\psi}
\newcommand{\SCI}{Ingelman_DIS95,Ingelman_LEPTO65,SCIa,SCIb,Ingelman_lishep98}
\newcommand{\POMPYT}{Ingelman_Brunia,Ingelman_Brunib,POMPYT}
\newcommand{\RAPGAP}{RAPGAP,RAPGAP206}
\newcommand{\EPJPSI}{epjpsi2,epjpsi3}
\newcommand{\BUCHMU}{Buchmueller_DIS95a,Buchmueller_DIS95b,Buchmuller_Hebecker_Mcdermott,Buchmueller_97a,Buchmueller_charm}
\newcommand{\pQCD}{Wusthoff,Diehl1,Diehl2,Bartels_dijet_ws,Bartels_jets,Bartels_asym}
\newcommand{\DGLAP}{DGLAPa,DGLAPb,DGLAPc,DGLAPd}
\newcommand{\JETSET}{Jetseta,Jetsetb,Jetsetc}
\newcommand{\kt}{kt_1,inv_kt,kt_2}
\input feynman
\bigphotons
\begin{document}
\thispagestyle{empty}
\noindent
DESY 98-131                           \hfill ISSN 0418--9833 \\
     \hfill LUNFD6/(NFFL-7158) 1998 \\
\begin{center}
  \begin{Large}
  \begin{bf}
{Monte Carlo Implementations of  Diffraction at HERA
\footnote{
To appear in {\em Proc. of the LISHEP workshop on diffractive physics},
  edited by A. Santoro (Rio de Janeiro, Brazil, Feb 16 - 20, 1998)
}}
  \end{bf}
  \end{Large}
\end{center}
\begin{center}
  \begin{large}
H.~Jung\\
    \vspace{0.5cm} 
    {\it Physics Department, Lund University, P.O. Box 118, 221 00 Lund,
Sweden}  
\end{large}
\\
\vspace*{1.cm}
  {\bf Abstract}
\end{center}
\begin{quotation}
\noindent
  {\bf Abstract}
\noindent
{\small
The Monte Carlo implementation of different approaches 
for diffractive scattering
in $e - p$ collisions (resolved $\PO$, pQCD, soft color interactions) 
is described, with emphasis on the construction of the
hadronic final state. Simple models for proton dissociation and 
exclusive vector meson production are described. 
A comparison of the different approaches is given. 
}
\end{quotation}
\section{Introduction}
Before the first observation of 
 events with large rapidity gaps in deep inelastic scattering  at HERA
 in 1993 \cite{ZEUS_diff,H1_diff},  diffractive reactions have been studied
 and were implemented into Monte Carlo programs:
 \par
The POMPYT~\cite{Ingelman_Brunic} Monte Carlo program is based on the
Ingelman-Schlein ansatz \cite{IS} assuming a partonic
structure of the pomeron $\PO$ and is suitable for a description of the
full hadronic final state in diffractive $e - p$ scattering.
\par
The DIFFVM~\cite{diffvm} Monte Carlo program
describes elastic vector-meson
production assuming the vector meson dominance model with a parameterization of
the vector-meson proton scattering cross section based on hadron hadron
scattering.
\par
The  EPJPSI~\cite{\EPJPSI} program, 
a Monte Carlo  for inelastic $\JP$ 
production also included a simple model for elastic $\JP$ production together
with a simulation of inelastic diffractive $\JP$ production, both based on
the Ingelman-Schlein ansatz \cite{IS}.
\par
Shortly after the observation of large rapidity gap events at HERA, 
the RAPGAP~\cite{\RAPGAP} Monte Carlo program,  and 
a improved version of the POMPYT~\cite{\POMPYT} Monte Carlo program 
became available. Like POMPYT,  RAPGAP was originally based
on the Ingelman-Schlein ansatz \cite{IS} and
 suitable for a description of the
full hadronic final state in diffractive $e - p$ scattering. 
With  the same ansatz, 
  diffractive final states were later also included
in the Monte Carlo program ARIADNE \cite{CDM,CDM_rapgap}.
\par
In 1995 a new approach to describe events with large rapidity gaps was proposed
by Edin, Ingelman and Rathsman~\cite{\SCI} and Buchm\"uller 
and Hebecker~\cite{\BUCHMU}.
 Independently both groups attempted to explain the
production of large rapidity gap events by interactions of the colored partons
from the hard interaction process 
with the color field of the proton. This interaction
was described  by the so-called soft color 
interactions (SCI)~\cite{\SCI} or
within a semi - classical approach~\cite{\BUCHMU}, respectively.
\par
In the meantime substantial progress in the theoretical understanding of
diffraction has been made, which enabled 
 a subset of diffractive scattering to be described 
in terms of perturbative QCD: 
Vector meson production (see for example \cite{Ryskin,Brodsky})
and the production of
exclusive high $p_T$ jets \cite{\pQCD} and charm \cite{Lotter_charm,Diehl_charm}
 (for a more complete list of references and a summary see \cite{diff_ws_sum}). 
The perturbative QCD approach for
high $p_T$ exclusive di-jet processes and heavy quark production 
has been implemented in the 
RAPGAP~\cite{RAPGAP206} Monte Carlo. Vector-meson production calculated in
perturbative QCD is implemented in the Monte Carlos 
DIPSI \cite{DIPSI},
RHODI \cite{RHODI}
and HITVM \cite{HITVM}. 
\par
In a completely different approach the pomeron $\PO$ is assumed to have 
direct couplings to quarks. This has been calculated and implemented in the 
Monte Carlo program VBLY~\cite{VBLY}.
\par
In the following I shall concentrate on the two Monte Carlo programs
RAPGAP~2.06~\cite{RAPGAP206} and LEPTO~6.5~\cite{Ingelman_LEPTO65}.
RAPGAP is discussed 
as a representative of multi - purpose Monte Carlo programs
 including a description
of 
 the resolved pomeron model according to the 
Ingelman-Schlein ansatz and the pQCD
description for diffractive $q\bar{q}$ production. Both the 
resolved pomeron model 
and the perturbative QCD description 
are successfully describing  present data on hadronic
energy flow and particle spectra 
as well as high $Q^2$ production of $J/\psi$ mesons.
LEPTO is  the only alternative model attempting to describe rapidity gap events
with the soft color interaction 
approach, without involving the concept of a Pomeron. 
A comparison of the different approaches (resolved pomeron, pQCD and  SCI) 
is made for a few experimental observables. 
\section{Kinematics and the total cross section}
In a diffractive 
process $ e + p \to e'+ p' + X$ where $p'$ represents the elastically
scattered proton or a low mass final state, and $X$ stands for the 
diffractive hadronic state,  
the cross section can be written as~\cite{Ingelman_Prytz}:    
\begin{equation}
\frac{d^4 \sigma (e p \to e' X p')}{dy\,dQ^2\,dx_{\PO}\,dt}
= \frac{4 \pi \alpha ^2}{y Q^4} 
   \left(   \left( 1 -  y + \frac{y^2}{2} \right) 
                F_2^{D(4)}(x,Q^2;x_{\PO},t)
                   -   \frac{y^2}{2} F_L^{D(4)}(x,Q^2;x_{\PO},t) \right)
\end{equation}
with $y=(q.p)/(e.p)$, $Q^2=-q^2=(e-e')^2$,$x_{\PO}= (q.\PO)/(q.p)$ 
and $t=(p-p')^2$ 
where $e$ ($e'$) are the four vectors of the incoming (scattered) electron,
the Bjorken $x$ variable $x= Q^2/(y\cdot s)$ with the total 
center of mass energy $s=(e+p)^2$,
$p$ ($p'$) are the four vectors of the incoming (scattered) proton,  
$q=e-e'$ is
the four vector of the exchanged photon and  
$\PO = p - p'$ corresponds to 
 the four vector of the pomeron. 
In analogy to Bjorken-$x$, one can define $\beta = x/x_{\PO}$.
In terms of experimental accessible quantities, these variables can be expressed
as:
\begin{eqnarray}
x_{\PO} & = &\frac{Q^2 + M_X^2}{Q^2 + W^2} \\
\beta & = & \frac{Q^2}{Q^2 + M_X^2} 
\end{eqnarray}
with $M_X$ being the invariant mass of the diffractive ($\gamma^* \PO$)
system and $W$ the mass of the $\gamma^* p$ system.
Independently of the underlying picture of diffraction
$x_{\PO}$ and $\beta$ can be defined.
However 
 the inclusive structure function $F_2^{D(4)}$ or equivalently the 
 $\gamma^* p$ cross section provides no direct  
information concerning the hadronic final state. In order to construct
a  Monte Carlo describing the hadronic final state,  
 the structure function $F_2^D$ has to be interpreted 
in terms of partonic subprocesses:
\begin{itemize}
\item {\bf Resolved pomeron a la Ingelman and Schlein} 

 In the Ingelman-Schlein model~\cite{IS}
 $F_2^{D(4)}$ can be written 
 as the product
of the probability of finding a pomeron, $f_{p\;\PO}$, in the proton
and the
structure function $F_2^{\PO}$ of the pomeron:
\begin{equation}
F_2^{D(4)}(\beta,Q^2;\xpom,t) = f_{p\;\PO}(\xpom,t)  F_2^{\PO}(\beta,Q^2)
\label{F2Ddef}
\end{equation}
In analogy to the quark - parton - model of the proton, 
$\beta$ can be interpreted as the momentum fraction of the total
pomeron momentum carried by the struck quark and
$ F_2^{\PO}(\beta,Q^2)$  can be described in terms of momentum weighted
quark density functions  in the pomeron.

\item {\bf pQCD calculation of diffraction} 

The pQCD calculation of diffraction is applicable mainly to  
exclusive high $p_T$
di-jet production, but in the model of \cite{Wusthoff} estimates on the
total inclusive diffractive cross section are given.
The calculation of diffractive di-jet production can be
performed using pQCD for large photon virtualities $Q^2$ and 
high $p_T$ of the 
$q (\bar{q})$ 
jets~\cite{\pQCD} or for heavy quarks \cite{Lotter_charm,Diehl_charm}.
\par
The process is mediated by two gluon exchange. Different
assumptions on the nature of the exchanged gluons 
can be
made: in \cite{Diehl1,Diehl2} 
the gluons are non perturbative, in \cite{Wusthoff}  
they are a hybrid of
non perturbative and perturbative ones and in 
\cite{Bartels_dijet_ws,Bartels_jets} 
they are
taken from a NLO parameterization of the proton structure 
function \cite{GRVa,GRVb}.
The cross section is essentially proportional to the gluon density 
squared of the proton: 
$\sigma \sim \left[\xpom G_p\left(\xpom,\mu^2 \right)\right]^2 $
with the scale $\mu^2 = p_T^2/(1-\beta)$.
In the case of heavy quarks the cross section is finite for all $p_T$, and
the scale is taken to be
$\mu^2 = (p_T^2 + m_f^2)/(1-\beta)$~\cite{Lotter_charm,Diehl_charm},
 with $m_f$ being the mass of the heavy quark.
\par
Due to the different gluon density parameterizations,
 different $\xpom$ dependencies of
the cross sections are expected and further discussed in
\cite{Bartels_dijet_ws,Bartels_jets}, 
where also numerical estimates are presented.

\item {\bf Semi-classical approach of Buchm\"uller, McDermott and Hebecker} 

Buchm\"uller et al.\ \cite{\BUCHMU}
attempt to describe $\gamma^* + p \to q + \bar{q} + p'$
and $\gamma^* + p \to q + \bar{q} + g + p'$ in a semi - classical approach where
the partons of the hard scattering subprocess
 interact with the color field of the proton.
 The cross section of the first process turns out
to be of similar structure as in the pQCD calculation of~\cite{Bartels_jets}
 and is proportional to a constant, which can be interpreted 
 in the semi-classical approach as the
 proton gluon density squared.
 The $q \bar{q} g$ process is
described with a usual boson gluon fusion subprocess,
 but involving 
 an
effective diffractive gluon density~\cite{Buchmueller_97a}:
\begin{equation}
\xpom g^D(\xpom,\beta) = \frac{C_1}{\beta \cdot C_g - \beta + 1} 
               \cdot  \frac{1}{\xpom}
\end{equation}
with $C_1$ and $C_g$ being free parameters.
Note that with $C_g=1$ the 
gluon density only depends on $\xpom$. Moreover in this approach only 
a $1/\xpom$ dependence appears,
 in contrast to other models, where   
a $1/\xpom^{1+\epsilon}$ is present. 
In order to account for non-zero $t$ the dipole form 
factor of the proton  is applied. 

\item {\bf Soft Color Interaction (SCI)} 

In this approach events with large rapidity gaps are produced by 
soft color interactions that change the color charge 
 of the  partons
originating from the hard interaction process~\cite{\SCI},
 before fragmentation. No pomeron is
explicitly introduced. All parameters in this model are determined by  
non-diffractive deep inelastic scattering, except the probability for a
 soft color interaction $R_{SCI}$.
\end{itemize}

\section{The partonic final state}
In this section 
I shall first describe the procedure to obtain a model of the
hadronic final state in diffraction using diffractive parton densities, which
I shall call the {\it resolved pomeron} model.
Here it is not necessary to assume factorization as in the 
Ingelman-Schlein ansatz.  
Then I shall describe the MC implementation of the perturbative QCD approach and
in which way it differs from 
 the resolved pomeron model.
At the end I shall discuss the soft color interaction (SCI) model.
\subsection{The partonic final state in the resolved pomeron
model}

In close analogy to inclusive $e-p$ scattering, where $F_2(x,Q^2)$ is given
by 
$ F_2(x,Q^2) = \sum_i e_i^2 \cdot x q_i(x,Q^2)$
with $e_i^2$ being the electric charge and $q_i (x,Q^2)$ the parton density
function of quark $i$, diffractive parton densities can be defined 
such, that the sum over $\xpom \cdot q^D_i(\beta,Q^2;\xpom,t)$ 
gives $F_2^{D(4)}$. 
With such a definition the hadronic
final state in diffractive processes can be constructed similarly to that of
non diffractive scattering, by the 0th order $\alpha_s$ process 
$\gamma ^* q \to q'$ (QPM) and the 1st order $\alpha_s$ processes 
$\gamma ^* q \to q g$ (QCDC) and $\gamma ^* g \to q \bar{q}$ (BGF).
In this procedure the incoming parton (a quark or a gluon) has
zero transverse momentum (except from a possible small intrinsic $k_T$), leaving
a remnant behind, which does not take part in the hard interaction and therefore
also has zero (or small) $k_T$. This is in contrast to the pQCD calculation of
\cite{\pQCD} for $\gamma ^* p \to q \bar{q} p'$, where both quark and 
anti-quark participate in the hard interaction and therefore have finite $k_T$,
without leaving a remnant behind. 
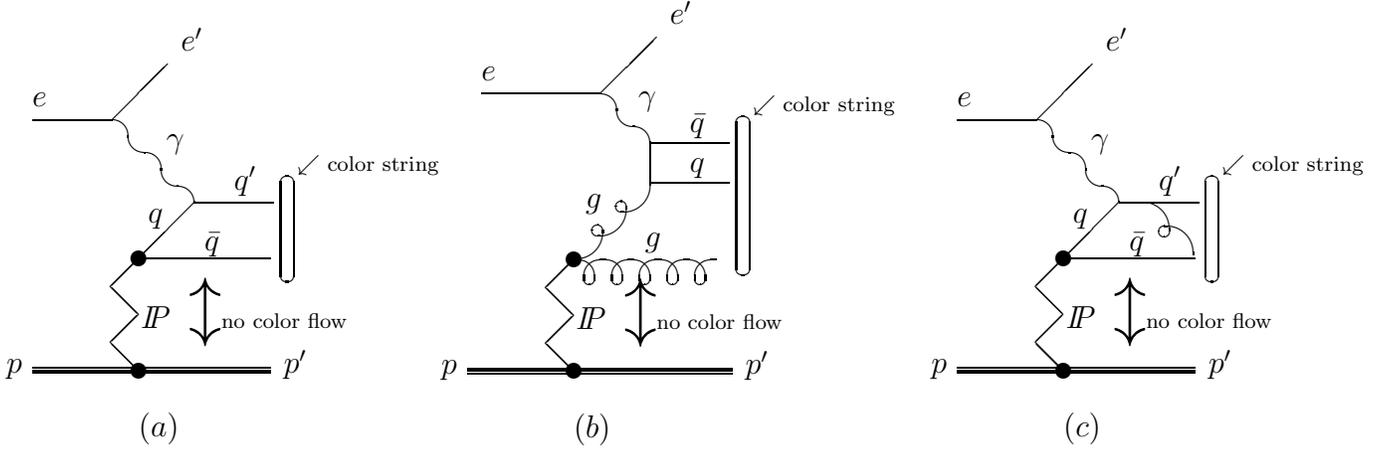
\begin{figure} [htb]
\begin{picture}(45000,20000) 
\drawline\fermion[\W\REG](3000,15000)[3000] 
\global\advance\pmidy by 500 
\put(\pbackx,\pmidy){$e$} 
\drawline\photon[\SE\REG](\fermionfrontx,\fermionfronty)[5] 
\global\advance\pmidx by 500 
\global\advance\pmidy by 500 
\put(\pmidx,\pmidy){$\gamma$} 
\drawline\fermion[\NE\REG](\fermionfrontx,\fermionfronty)[3000] 
\global\advance\pbackx by 500 
\global\advance\pbacky by 500 
\put(\pbackx,\pbacky){$e'$} 
\drawline\fermion[\E\REG](\photonbackx,\photonbacky)[3000] 
\global\advance\pmidy by 500 
\global\advance\pbackx by 500 
\global\Xfive = \pbackx 
\global\advance\pbacky by - 1000 
\global\Yfive = \pbacky 
\put(\pmidx,\pmidy){$q'$} 
\drawline\fermion[\SW\REG](\photonbackx,\photonbacky)[3000] 
\global\advance\pmidx by -700 
\global\advance\pmidy by 300 
\put(\pmidx,\pmidy){$q$} 
\put(\Xfive,\Yfive){\oval(500,4000)} 
\global\advance\Yfive by + 2200 
\put(\Xfive,\Yfive){ \scriptsize $\swarrow$ color string } 
\put(\pbackx,\pbacky){\circle*{600}} 
\global\Xone = \pbackx 
\global\Yone = \pbacky 
\drawline\fermion[\E\REG](\Xone,\Yone)[5000] 
\global\advance\pmidy by 300 
\put(\pmidx,\pmidy) {$\bar{q}$} 
\drawline\fermion[\SW\REG](\Xone,\Yone)[1500] 
\drawline\fermion[\SE\REG](\pbackx,\pbacky)[1500] 
\drawline\fermion[\SW\REG](\pbackx,\pbacky)[1500] 
\global\advance\pmidx by + 600 
\put(\pmidx,\pmidy){$\PO \;\;${\Huge $\updownarrow$}\scriptsize no color flow} 
\drawline\fermion[\SE\REG](\pbackx,\pbacky)[1500] 
%
%
\put(\pbackx,\pbacky){\circle*{600}} 
\global\advance\pbacky by - 2500 
\put(\pbackx,\pbacky){$ (a) $} 
\global\advance\pbacky by + 2500 
\global\advance\pbacky by -100 
\multiput(\pbackx,\pbacky)(0,100){3}{\line(-1,0){4000}}   
\global\Xthree = \pbackx 
\global\Ythree = \pbacky 
\global\advance\Xthree by - 5000 
\put(\Xthree,\Ythree){$p$} 
\multiput(\pbackx,\pbacky)(0,100){3}{\line(1,0){5000}} 
\global\advance\pbackx by + 5500 
\put(\pbackx,\pbacky){$p'$} 
\drawline\fermion[\W\REG](21500,16000)[4500] 
\global\advance\pmidy by 500 
\put(\pbackx,\pmidy){$e$} 
\drawline\photon[\SE\REG](\fermionfrontx,\fermionfronty)[3] 
\global\advance\pmidx by 500 
\global\advance\pmidy by 500 
\put(\pmidx,\pmidy){$\gamma$} 
\drawline\fermion[\NE\REG](\fermionfrontx,\fermionfronty)[3000] 
\global\advance\pbackx by 500 
\global\advance\pbacky by 500 
\put(\pbackx,\pbacky){$e'$} 
\drawline\fermion[\E\REG](\photonbackx,\photonbacky)[3000] 
\global\advance\pmidy by 400 
\put(\pmidx,\pmidy){$\bar{q}$} 
\drawline\fermion[\S\REG](\photonbackx,\photonbacky)[1500] 
\drawline\fermion[\E\REG](\fermionbackx,\fermionbacky)[3000] 
\global\advance\pmidy by 400 
\put(\pmidx,\pmidy){$q$} 
\global\advance\pbackx by 500 
\global\Xfive = \pbackx 
\global\advance\pbacky by - 500 
\global\Yfive = \pbacky 
\drawline\gluon[\SW\REG](\fermionfrontx,\fermionfronty)[2] 
\global\advance\pmidx by -1000 
\global\advance\pmidy by 500 
\put(\pmidx,\pmidy){$g$} 
\put(\Xfive,\Yfive){\oval(500,6000)} 
\global\advance\Yfive by + 2200 
\put(\pbackx,\pbacky){\circle*{600}} 
\global\advance\Yfive by + 1000 
\put(\Xfive,\Yfive){ \scriptsize $\swarrow$ color string } 
\global\Xone = \pbackx 
\global\Yone = \pbacky 
\drawline\gluon[\E\REG](\Xone,\Yone)[5] 
\global\advance\pmidy by 500 
\put(\pmidx,\pmidy) {$g$} 
\drawline\fermion[\SW\REG](\Xone,\Yone)[1500] 
\drawline\fermion[\SE\REG](\pbackx,\pbacky)[1500] 
\drawline\fermion[\SW\REG](\pbackx,\pbacky)[1500] 
\global\advance\pmidx by + 600 
\put(\pmidx,\pmidy){$\PO \;\;${\Huge $\updownarrow$}\scriptsize no color flow} 
\drawline\fermion[\SE\REG](\pbackx,\pbacky)[1500] 
%
%
\put(\pbackx,\pbacky){\circle*{600}} 
\global\advance\pbacky by - 2500 
\put(\pbackx,\pbacky){$ (b) $} 
\global\advance\pbacky by + 2500 
\global\advance\pbacky by -100 
\multiput(\pbackx,\pbacky)(0,100){3}{\line(-1,0){4000}} 
\global\Xthree = \pbackx 
\global\Ythree = \pbacky 
\global\advance\Xthree by - 5000 
\put(\Xthree,\Ythree){$p$} 
\multiput(\pbackx,\pbacky)(0,100){3}{\line(1,0){6000}} 
\global\advance\pbackx by + 6500 
\put(\pbackx,\pbacky){$p'$} 
\drawline\fermion[\W\REG](38000,15000)[3000] 
\global\advance\pmidy by 500 
\put(\pbackx,\pmidy){$e$} 
\drawline\photon[\SE\REG](\fermionfrontx,\fermionfronty)[5] 
\global\advance\pmidx by 500 
\global\advance\pmidy by 500 
\put(\pmidx,\pmidy){$\gamma$} 
\drawline\fermion[\NE\REG](\fermionfrontx,\fermionfronty)[3000] 
\global\advance\pbackx by 500 
\global\advance\pbacky by 500 
\put(\pbackx,\pbacky){$e'$} 
\drawline\fermion[\E\REG](\photonbackx,\photonbacky)[3000] 
\global\advance\pmidy by 500 
\global\advance\pbackx by 500 
\global\Xfive = \pbackx 
\global\advance\pbacky by - 1000 
\global\Yfive = \pbacky 
\put(\pmidx,\pmidy){$q'$}
\global\advance\pmidx by -700 
\drawline\gluon[\SE\REG](\pmidx,\photonbacky)[1] 
\global\advance\pmidx by +700 
\drawline\fermion[\SW\REG](\photonbackx,\photonbacky)[3000] 
\global\advance\pmidx by -700 
\global\advance\pmidy by 300 
\put(\pmidx,\pmidy){$q$} 
\put(\Xfive,\Yfive){\oval(500,4000)} 
\global\advance\Yfive by + 2200 
\put(\Xfive,\Yfive){ \scriptsize $\swarrow$ color string } 
\put(\pbackx,\pbacky){\circle*{600}} 
\global\Xone = \pbackx 
\global\Yone = \pbacky 
\drawline\fermion[\E\REG](\Xone,\Yone)[5000] 
\global\advance\pmidy by 300 
\put(\pmidx,\pmidy) {$\bar{q}$} 
\drawline\fermion[\SW\REG](\Xone,\Yone)[1500] 
\drawline\fermion[\SE\REG](\pbackx,\pbacky)[1500] 
\drawline\fermion[\SW\REG](\pbackx,\pbacky)[1500] 
\global\advance\pmidx by + 600 
\put(\pmidx,\pmidy){$\PO \;\;${\Huge $\updownarrow$}\scriptsize no color flow} 
\drawline\fermion[\SE\REG](\pbackx,\pbacky)[1500] 
%
%
\put(\pbackx,\pbacky){\circle*{600}} 
\global\advance\pbacky by - 2500 
\put(\pbackx,\pbacky){$ (c) $} 
\global\advance\pbacky by + 2500 
\global\advance\pbacky by -100 
\multiput(\pbackx,\pbacky)(0,100){3}{\line(-1,0){4000}} 
\global\Xthree = \pbackx 
\global\Ythree = \pbacky 
\global\advance\Xthree by - 5000 
\put(\Xthree,\Ythree){$p$} 
\multiput(\pbackx,\pbacky)(0,100){3}{\line(1,0){5000}} 
\global\advance\pbackx by + 5500 
\put(\pbackx,\pbacky){$p'$} 
\end{picture} 
\caption{Basic processes for inelastic diffractive 
lepton nucleon scattering.
Indicated are the color strings and the pomeron remnant.
$a$. shows the QPM process (0th order $\alpha_s$) for quark scattering.
$b$. shows the
$O(\alpha _{em}\alpha_s)$
for photon gluon fusion (the crossed diagram is not shown).
 The pomeron remnant is a color octet gluon.
 In $c$.  the
$O(\alpha _{em}\alpha_s)$  
QCD Compton process (the crossed diagram is not shown) is shown.
 The pomeron remnant is the same as in $a$.
\label{difmc}}
\end{figure}
\par
 With the knowledge of $\xpom \cdot q^D_i(\beta,Q^2;\xpom,t)$
and $\xpom \cdot g^D_i(\beta,Q^2;\xpom,t)$
 the total cross section can be described in 
terms of scattering off a virtual photon 
on a quark or anti-quark. However this quark may have
been originated from another parton, producing
 a different 
hadronic final state. The process where an initial parton 
carrying a 
momentum fraction $x_i$, splits into other partons which then undergo 
hard scattering with the photon, can be  simulated in QCD parton showers
 based on the 
leading log DGLAP \cite{\DGLAP} splitting functions
in leading order $\alpha_s$. 
\par
A more detailed simulation of 
leading order $\alpha_s$ processes like $\gamma^* g \to q \bar{q}$ 
(BGF, Fig.~\ref{difmc}b.)
and $\gamma^* q \to q g$ (QCD - Compton, Fig.~\ref{difmc}c.)
 is obtained when 
the exact $O(\alpha_s)$ QCD matrix elements for these processes are included.
\par
 The decision whether to generate a QPM or a 1st order 
$\alpha_s$ event, is based on their relative cross sections
at a given $x$ and $Q^2$. 
Technically for each event the cross section
for BGF light quarks, BGF heavy quarks and QCD - Compton has to be 
obtained from a  numerical integration including the proper parton densities.
If the scale chosen for
 $\alpha_s$ and the parton densities  
is either
$Q^2$ or the invariant mass squared of the two hard partons, 
 $\hat{s}$,  
then the matrix elements can be integrated 
analytically over one degree of freedom
 leaving only a one dimensional numerical integration to be done.
If, however, the scale is $p_T^2$ (or any function of it) then $\alpha_s$ and
the parton densities cannot be factorized, and a time consuming 
two dimensional numerical integration has to be performed.
As an alternative the QCD probabilities can be calculated once and stored in
a grid.
 This approach is faster but less accurate. Both options for obtaining the
 $O(\alpha_s)$ QCD probabilities are implemented in RAPGAP and LEPTO, where for
 the latter 
  only the scale $Q^2$ is implemented.
\par
In order to avoid divergences in the matrix elements for massless quarks
a cutoff in $p_T^{cut}$ (or in other variables like
$\hat{s}$ and $z$  or $\hat{t}$) has to be specified.
 The minimum $p_T^{cut}$ is at least restricted by the 
requirement that the sum 
of the order $\alpha_s$ processes has to be smaller 
or equal to the total cross section given by $F_2$ (or $F_2^{D(4)}$).
Note that if a too small $p_T^{cut}$ is used, the prediction from the 
1st order $\alpha_s$ matrix elements might be unreliable since 
$\alpha_s$ might be too large to justify the use of pQCD.
\par
Having thus constructed the hard scattering subprocess up to order $\alpha_s$, 
higher order corrections may be simulated by initial and final state parton
showers. Because of the strong ordering of virtualities in a DGLAP evolution,
the virtuality in the parton showers is restricted by the cutoff $p_T^{cut}$ for
 the
$O(\alpha_s)$ matrix elements, since partons having larger $p_T>p_T^{cut}$
are already generated by the matrix element processes.
\par
This ends the construction of the hard partonic final state, 
leaving the construction
of the final sate proton or proton dissociation to be done. The hadronic 
final state is then constructed by handing the partons over to a fragmentation
program like JETSET~\cite{\JETSET}.

\subsection{Perturbative QCD approach to diffraction}
The cross section for $e p \to e' q\bar{q} p$ has been calculated in
\cite{\pQCD}. For the production of light quarks a $p_T^{cut}$ has to be applied,
to regulate the collinear divergence of the matrix element. In the case of heavy
quark production \cite{Lotter_charm,Diehl_charm} the cross section is finite
even for small $p_T$ because of the heavy quark mass.
\par
Since both the quark and the 
anti-quark participate in the hard interaction, they both
receive the same transverse momentum in the $\gamma^* \PO$ system, without
producing  a remnant, 
and both final state partons are allowed to further radiate
partons in the final state parton shower. This has to 
be contrasted to the
resolved pomeron model, where also a $q\bar{q}$ final state appears in a QPM 
process, but there the  quarks have vanishing transverse momentum 
(except from a small intrinsic $p_T$) in the $\gamma^* \PO$ center of mass
system and a pomeron remnant is present. 
 Another striking feature of the
exclusive $q\bar{q}$ production in pQCD 
 is a very special azimuthal asymmetry between jet and
the lepton plane in the $\gamma^* \PO$ CMS.
\par
These processes have been implemented in RAPGAP, allowing different 
 parameterizations of the gluon density of the proton to be used. 
\subsection{Soft color interaction}
A detailed description of the soft color interaction model can be found 
in~\cite{\SCI}. Here only the main characteristics are 
presented.
\par
 In non - diffractive scattering at small $x$ the most important process is 
 boson - gluon fusion ($\gamma ^* g \to q \bar{q}$). 
 After the hard
scattering process and the initial and final state QCD radiation
took place, the partons of the hard scattering process 
 travel through the color field of the proton and there is a
certain probability for soft color interactions, which can change the color 
 structure without changing the kinematics of the process. The 
 probability for soft color interactions $R_{SCI}$ is the only free 
 parameter in this model, where $R_{SCI}$ may be seen as
 a value of the strong coupling constant
 $ \frac{\alpha_s}{\pi}\sim 0.2$ at a scale of 0.5  GeV, 
 which is 
 representative for the region below the perturbative cutoff.
  Large rapidity gap
 events can be produced since it can happen that the hard scattering 
 subprocess becomes disconnected in color space 
 from partons of the QCD cascade and
 the proton remnant. 
 For large rapidity gap events a
  $\sim 1/M_X^2$ distribution
  is generated from the $\sim 1/\hat{s}$ 
 dependence of the BGF process. Also the $t$ distribution follows in 
 general an exponential form, determined by the Gaussian width of the
 distribution of the
 intrinsic transverse momentum of the partons 
 in the proton (for details see \cite{\SCI}).
 \par
 The SCI model allows a smooth transition between non-diffractive to 
 diffractive scattering, including also a simulation of Reggeon and $\pi$
 exchange processes, without introducing these exchanges explicitly.
\section{A simple model for proton dissociation}
In this section a simple model for proton dissociation implemented in RAPGAP and
based on ideas from PYTHIA~\cite{Jetsetc} is described.
When a proton dissociates, it can split into a quark $q_p$ and di-quark $di-q_p$
system. The pomeron is assumed to couple to a single quark 
$q_p$ only, 
and therefore the outgoing quark $q'_p$ carries 
 all of the momentum transfer $t$ resulting in a finite transverse momentum. 
The quark to which the pomeron 
couples carries a momentum fraction $\chi$ of the protons initial
momentum. The momenta of the initial quark $q_p$ and the di-quark
$di-q_p$ are:
\begin{eqnarray}
q_p & \simeq & \chi p \\
di-q_p & \simeq & (1 - \chi) p \\
q'_p & = & q_p - \PO \simeq  \chi p - \PO
\end{eqnarray}
where  
$ \chi = \frac{q.q_p}{q.p}$ with
$q$ ($q_p$, $p$ and $\PO$) being the photon (quark, proton and 
pomeron) momentum, respectively.
 In addition the quark and
di-quark can receive a primordial $p_{\perp}$ according to a Gaussian 
distribution. 
\par
The momentum fraction $\chi$ can be estimated within the 
resolved pomeron model:
\begin{equation}
\chi = \frac{\xpom}{\beta'}
\end{equation}
with $\beta' = \frac{q.\PO}{q.q_p}$ 
being the fraction of the pomeron momentum carried by the quark $q_p$
and  $x_{\PO} = \frac{q.\PO}{q.p}$.  $\beta '$ is defined similarly to 
$\beta$, which is  used in the structure function $F_2^{\PO}$.
The value of $\xpom$ is already known from the interaction $\gamma \PO$
and $\beta'$ can be generated according to the quark density of the pomeron.
\par
Instead of using explicitly the parameterization of parton densities in the
pomeron, which have been obtained from $F_2^D$, a more simple ansatz is chosen,
since the scale, at which proton dissociation happens is too small 
(typically of the order of $|t| \ll 1$ GeV$^2$) to be used in parton density
function parameterizations. 
 For  $\beta'$ different, alternative probability functions $P_i$
are available: 
\begin{eqnarray}
P_1(\beta')  = & 2 (1 - \beta')  \\ 
P_2(\beta')  =  & (a+1) (1 - \beta')^a  \\
P_3(\beta')  =  & \frac{N}{ \beta'\left(1 - \frac{1}{\beta'} - 
          \frac{c}{(1 - \beta')}\right) ^2}                
\end{eqnarray}
 with $a$ chosen such that 
$<\beta'> = 1/(a + 2)$ 
and  $c$ determined by the ratio of masses of the
remnant quark and di-quark system. The first option   
corresponds to a hard quark density of the pomeron. 
These parameterizations are actually used as 
the probability functions even in non-diffractive scattering and have been
taken from  LEPTO~\cite{MEPS}.

\section{Vector-meson production}
In RAPGAP
vector meson production is naturally included  in diffractive scattering.
This is easiest seen for $J/\psi$ production. Suppose we have a system of
a $c \bar{c}$ quark pair, plus possibly additional gluons in the final state.
If the invariant mass 
$4\cdot m_c^2 <m_x^2=(q + p_{\pom})^2 < 4 \cdot m^2_{D^0}$ 
then only $J/\psi$'s can be produced ($\eta_c$ production is not possible
because of spin constraints). 
Technically vector-mesons are produced in the fragmentation
program JETSET \cite{\JETSET}, with the additional restriction that only 
 spin 1 mesons are generated.
However there is an uncertainty in the normalization of the cross section,
depending on the actual value of $m_c$ used. 
 The production of the
light vector-mesons $\rho$, $\omega$ and $\phi$ proceeds in a similar way to
that of the $J/\psi$, with the ratio $\rho:\omega=
9:1$ being fixed and $\phi$ vector mesons being only produced from $s$ quarks. 
\begin{figure}[htb]
\begin{center}
\epsfig{figure=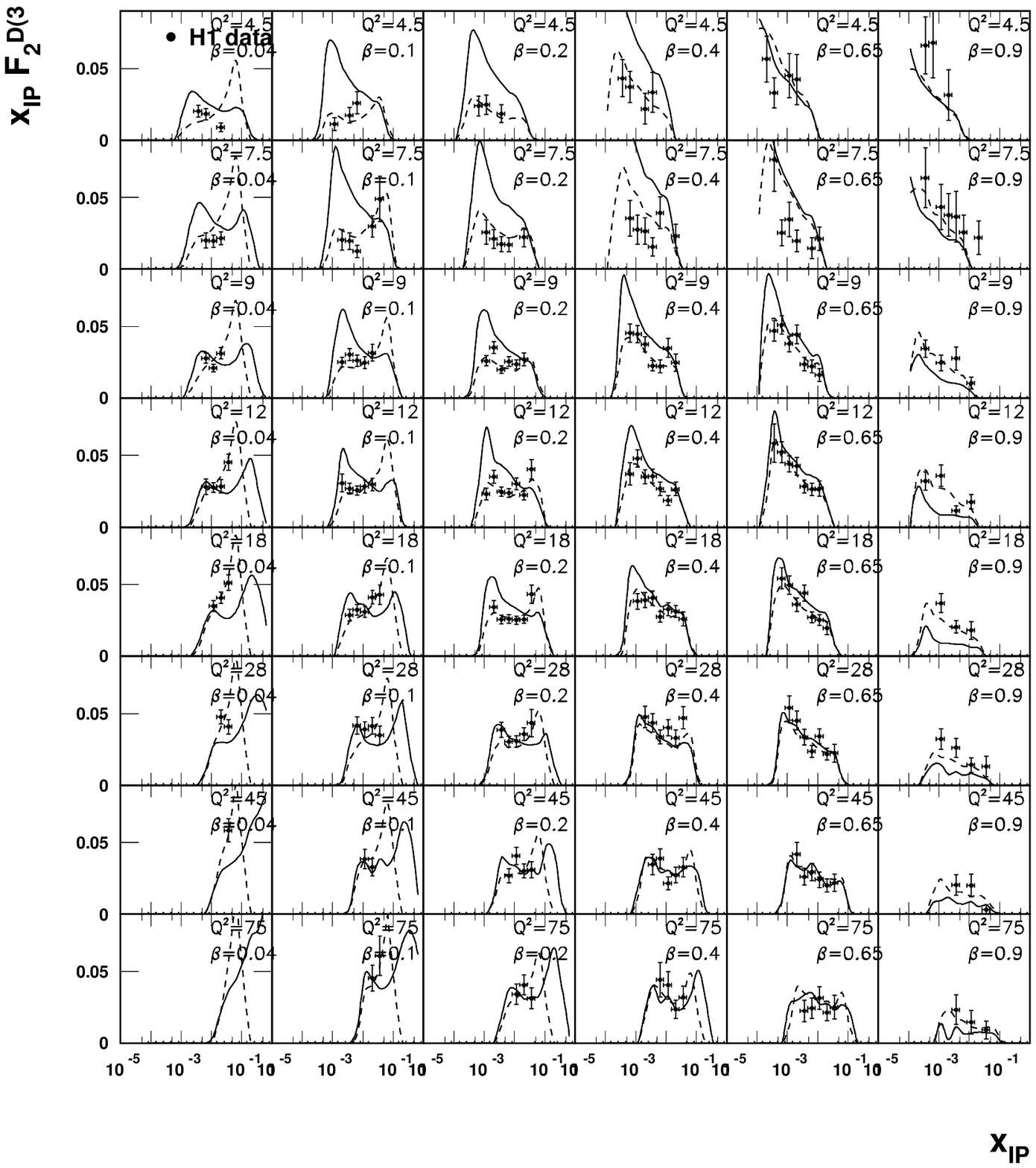,width=17cm,height=19cm}
\end{center}
\caption{The diffractive structure function 
$\xpom \cdot F_2^{D(3)}$~\protect\cite{H1_F2D3_97}.
The solid line shows the prediction  of the soft color interaction model
LEPTO 6.5 with $R_{SCI}=0.5$. For comparison also the result from the resolved
pomeron model is shown, using the $Q^2$ evolved parton densities from a fit to
 $F_2^{D(3)}$.
\label{F2D3}}
\end{figure}
\par
Exclusive vector meson production implies certain restrictions on the
kinematic variables $\xpom$ and $\beta$:
\begin{eqnarray}
\xpom = & \frac{q.p_{\pom}}{q.P}   =  
        \frac{Q^2 + M_X ^2}{Q^2 + W^2} \\
\beta = & \frac{Q^2}{2\cdot q.p_{\pom}}  =
        \frac{Q^2}{Q^2 + M_X^2} 
\end{eqnarray}	  
Thus for $M_X=m_{VM}$ and  fixed $W$ 
the variables $\xpom$ and $\beta$ depend only on
$Q^2$. Thus varying $Q^2$ implies varying 
$\xpom$ and  $\beta$. The effective $Q^2$ dependence
of the $\gamma ^* p \rightarrow \mbox{VM} p$ cross section is stronger 
($\simeq 1/Q^4$), than
 expected from $\gamma^* p \to X$ ($\simeq 1/Q^2$).

\section{Model Comparisons}
In this section I shall compare the different approaches to describe
diffraction and compare them  with data, where available.
\subsection{Inclusive structure function $F_2^D$}
 In Fig.~\ref{F2D3} the prediction for the 
 diffractive structure function $F_2^{D(3)}$
 of the SCI model (using $R_{SCI}=0.5$)  as implemented in
 LEPTO 6.5~\cite{Ingelman_LEPTO65} is compared to the measurement of 
 H1~\cite{H1_F2D3_97}. It is remarkable that this model, with essentially only
 one free parameter, $R_{SCI}$ is able to describe the general trend of the
 data. However in the low $Q^2$ and low $\beta$ region this model overshoots the
 data.
 For comparison  the results from the resolved pomeron model
 are also shown in Fig.~\ref{F2D3}. The fact that this 
 model describes the data is not surprising, since 
 the $Q^2$ evolved parton densities obtained from a fit to
 $F_2^{D(3)}$~\cite{H1_F2D3_97}  are used.
 \par
 The total cross section of the 
 perturbative QCD approach is not shown, since it depends on 
    the $p_T$ cutoff needed
 for the light quark contribution.
\subsection{Proton Dissociation}
\begin{figure}[htb]
\begin{center}
\epsfig{figure=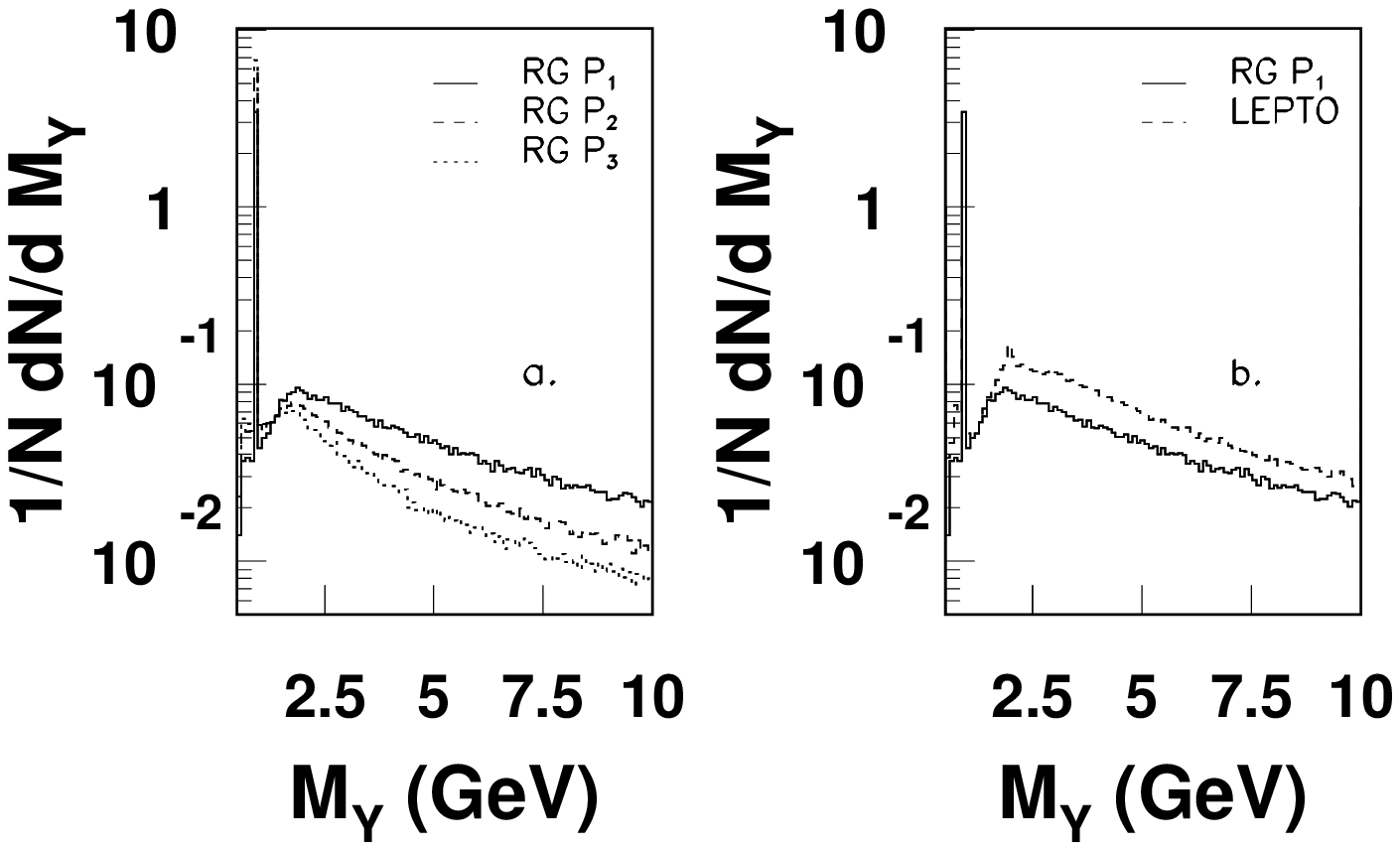,width=16cm,height=11cm}
\end{center}
\caption{The spectrum $M_Y$ for $e-p$ events,
 with $x_{\PO} < 0.1$ and $|t|<1$ GeV$^2$ in the range 
 $0.01<y<0.6$, $5<Q^2<80$. 
In $a.$ the RAPGAP predictions for different parameterization of the
quark splitting functions $P_i(\beta ')$, as described in the text, are shown.
In $b.$ the solid line shows the prediction from RAPGAP with $P_1(\beta') $ and
the dashed line shows the prediction from LEPTO 6.5 with $R_{SCI}=0.5$.
\label{pdiss}}
\end{figure}

Experimentally proton dissociation can be studied, when the system
$M_X$, associated with the $\gamma$ vertex, and the system $M_Y$, associated
with the proton vertex, can be separated. This can be achieved by searching for
the largest rapidity gap in an event
 following the procedure of 
 \cite{H1_F2D3_97}. The  mass distribution $M_Y$ of the $p$ - dissociative
system in $e-p$ diffractive events obtained from the RAPGAP Monte Carlo 
is shown in 
Fig.~\ref{pdiss}$a$. 
 The difference between the different
options for $P(\beta')$ is shown. In Fig.~\ref{pdiss}$b$. a comparison between
RAPGAP and LEPTO is shown.
At small masses $M_Y$ ($M_Y<10 $ GeV)
a $\sim 1/M_Y$ spectrum is obtained both for RAPGAP and LEPTO and only at 
larger masses $M_Y>10 $ GeV (not shown in Fig.~\ref{pdiss})
a typical $\sim 1/M_Y^2$ dependence 
is observed.
Even with proton dissociation switched on, a proton will emerge after
fragmentation when 
the momentum transfer is small 
and the mass of the $q$ - $ di-q$ system remains below
the threshold for multi-particle production.
\par
The SCI model automatically also simulates
 proton dissociation
since the color flow between the hard scattering and the proton remnant can be 
broken anywhere in rapidity.
 In Fig.~\ref{pdiss} the $M_Y$ spectrum obtained from
 LEPTO 6.5 is shown. Also in the SCI model a general $\sim 1/M_Y$ 
 spectrum is obtained, however the ratio of proton - elastic to 
 proton  dissociative events is different to that obtained from RAPGAP.
\subsection{Hadronic energy flow}
\begin{figure}[htb]
\begin{center}
\epsfig{figure=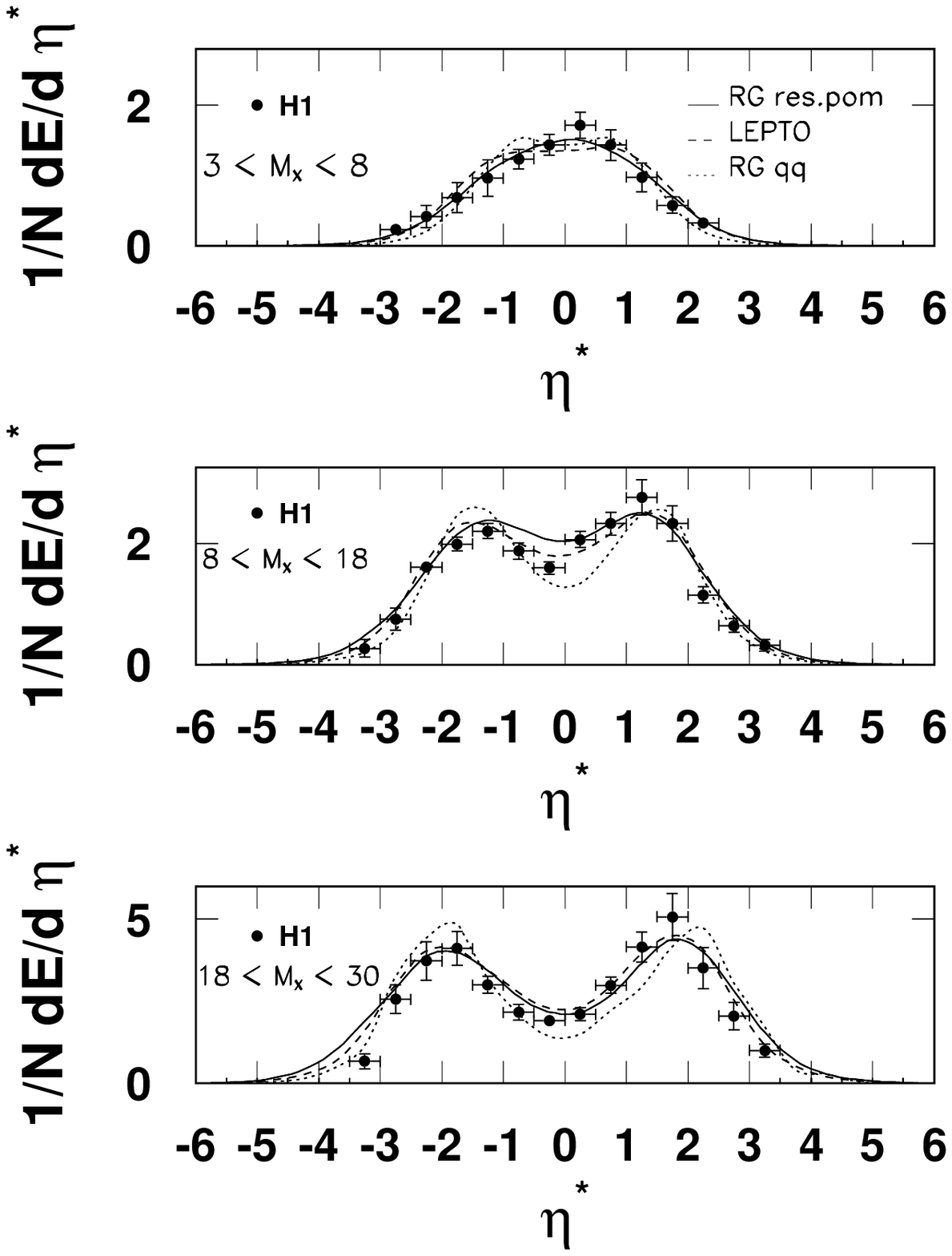,width=17cm,height=19cm}
\end{center}
\caption{The energy flow in the $\gamma^* \PO$ system $M_X$ as a function
of the pseudo rapidity in 3 different regions of $M_X$ (in GeV) indicated
 as measured by  
H1~\protect\cite{H1_eflow_diff}. The solid line is the prediction of RAPGAP
in the resolved pomeron mode using fit 2 of the H1 parameterization of 
$F_2^{D(3)}$ \protect\cite{H1_F2D3_97}. The dashed curve is the prediction
of LEPTO~6.5 using $R_{SCI} = 0.5$ with the GRV structure function of the
proton.
The dotted curve shows the prediction of the perturbative QCD calculation
as implemented in RAPGAP.
\label{had}}
\end{figure}

In Fig.~\ref{had} the energyflow in diffractive scattering 
predicted by the various models 
is compared to the measurement of H1~\cite{H1_eflow_diff}.
 The prediction of the
 SCI model (dashed line in Fig.~\ref{had})
 is very close to the prediction of the resolved pomeron model in
 RAPGAP (solid line), and both give a good description of the data.
 The SCI model is rather successful in describing properties of
 the hadronic final state in diffraction without involving the concept of a
 pomeron.
 \par
Also shown in
 Fig.~\ref{had} is the energy flow predicted from 
 the perturbative QCD calculation (dotted line) as implemented
in RAPGAP. 
The gluon density was taken from the GRV HO parameterization and   
$p_T> 1$ GeV was required for
the final state quarks. Both final state quarks
were allowed to emit further  QCD radiation. There is a remarkable 
agreement between this prediction and the data, considering that it has no
free parameter left.
In the lowest $M_X$ bin at central rapidities ($\eta^* \sim 0$) a small dip
is observed, which is understood 
as a consequence of  the finite $p_T$ cutoff for the
final state quarks.
 One has to note that no free parameter is left in this 
calculation. It is a great success that the 
energy flow can be understood in terms of perturbative QCD. 
\par
From the comparison of the different approaches to describe diffraction
with data on the hadronic energy flow, no
final conclusion on the underlying physics process can yet 
be made. All the very
different approaches describe the available data reasonably well. Different
and more sensitive measurements are obviously needed to differentiate between the
various approaches.
\subsection{$\phi$ asymmetries of jets and heavy quarks}
The very specific signature of the perturbative QCD calculation becomes
apparent only, when more detailed final state properties are considered.  
One of the main differences of the perturbative QCD calculation to both the
resolved pomeron model and as well as to the SCI model is the absence of 
a pomeron remnant.
This a signature can be investigated with  exclusive high $p_T$ di - jet events.
In that case the invariant mass of the jet - jet system 
$\hat{s}_{jj}$ is identical to the total invariant mass of the diffractive 
system $M_X^2$.
However depending on the
jet algorithm used to identify the high $p_T$ jets, a certain fraction of 
hadronic energy might not be associated to the jets, which might cause a problem
for the identification of exclusive di - jet events. 
\par
\begin{figure}[htb]
\begin{center}
\epsfig{figure=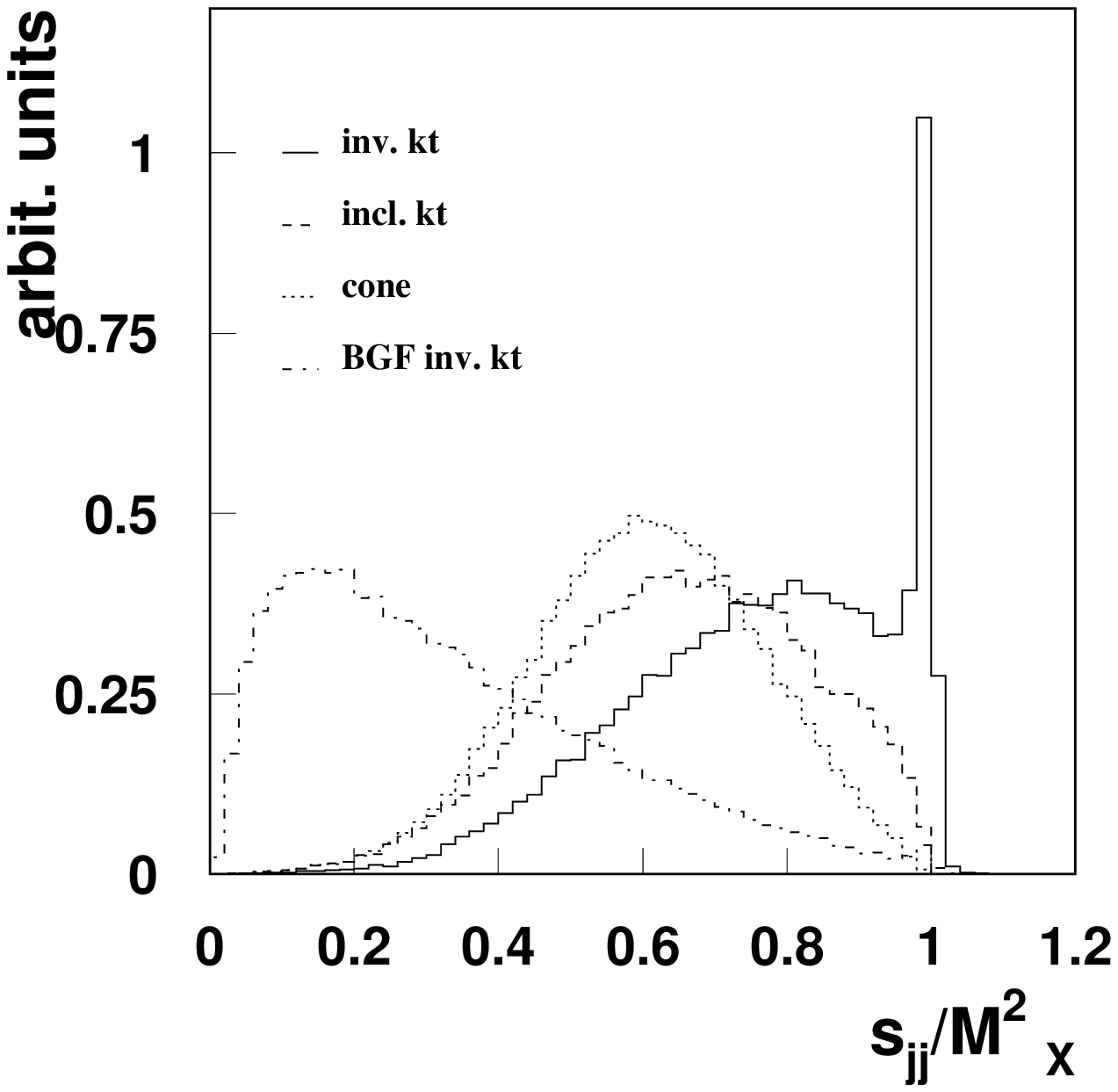,width=13cm,height=13cm}
\end{center}
\caption{The ratio $R_{jj} = \hat{s}_{jj} / M_X^2$
of the invariant mass squared of the jet-jet system to the total diffractive mass
$M_X^2$ for exclusive di-jet production according to the pQCD calculation.
The solid line is obtained using the 
$k_T$ jet algorithm with a  $y_{cut}$ such that two hard
 jets with $k_T > 2$ GeV 
 and a possible remnant jet are reconstructed in the Breit frame
 (labeled inv. $k_T$), the
dashed line with the inclusive $k_T$ jet algorithm and the dotted line with the
cone jet algorithm.  For comparison the distribution obtained from the resolved
$\PO$ model using BGF processes are shown with the dashed-dotted line. Please
note that the normalization of the distributions is arbitrary.
\label{dijets}}
\end{figure}

For a study of the effects of different jet algorithms, the 
following kinematic cuts are applied: $0.1 < y < 0.7$, $5 < Q^2< 80$ GeV$^2$
$\xpom < 0.05$ and $p_T^{jet} > 2$ GeV.
 Using the $k_T$ jet algorithm~\cite{\kt} with a  $y_{cut}$ such that two hard
 jets with $k_T > 2$ GeV 
 and a possible remnant jet are reconstructed in the Breit frame 
 (labeled invariant $k_t$),
 a distribution of
$R_{jj} = \hat{s}_{jj} / M_X^2$ is obtained, as shown in Fig.~\ref{dijets} 
with the solid line histogram.
For an ideal jet reconstruction $R_{jj} =1 $ is expected. 
However one sees large
effects due to the jet reconstruction.
The results using  different jet algorithms
like  cone type jet algorithm~\cite{pxcone}  or the inclusive $k_T$ 
algorithm~\cite{\kt} are also shown in 
Fig.~\ref{dijets}. For comparison also the distribution obtained from a BGF type
process in the resolved pomeron model is shown.
Exclusive di - jet events are best identified using the invariant $k_t$
algorithm.
\par
\begin{figure}[htb]
\begin{center}
\epsfig{figure=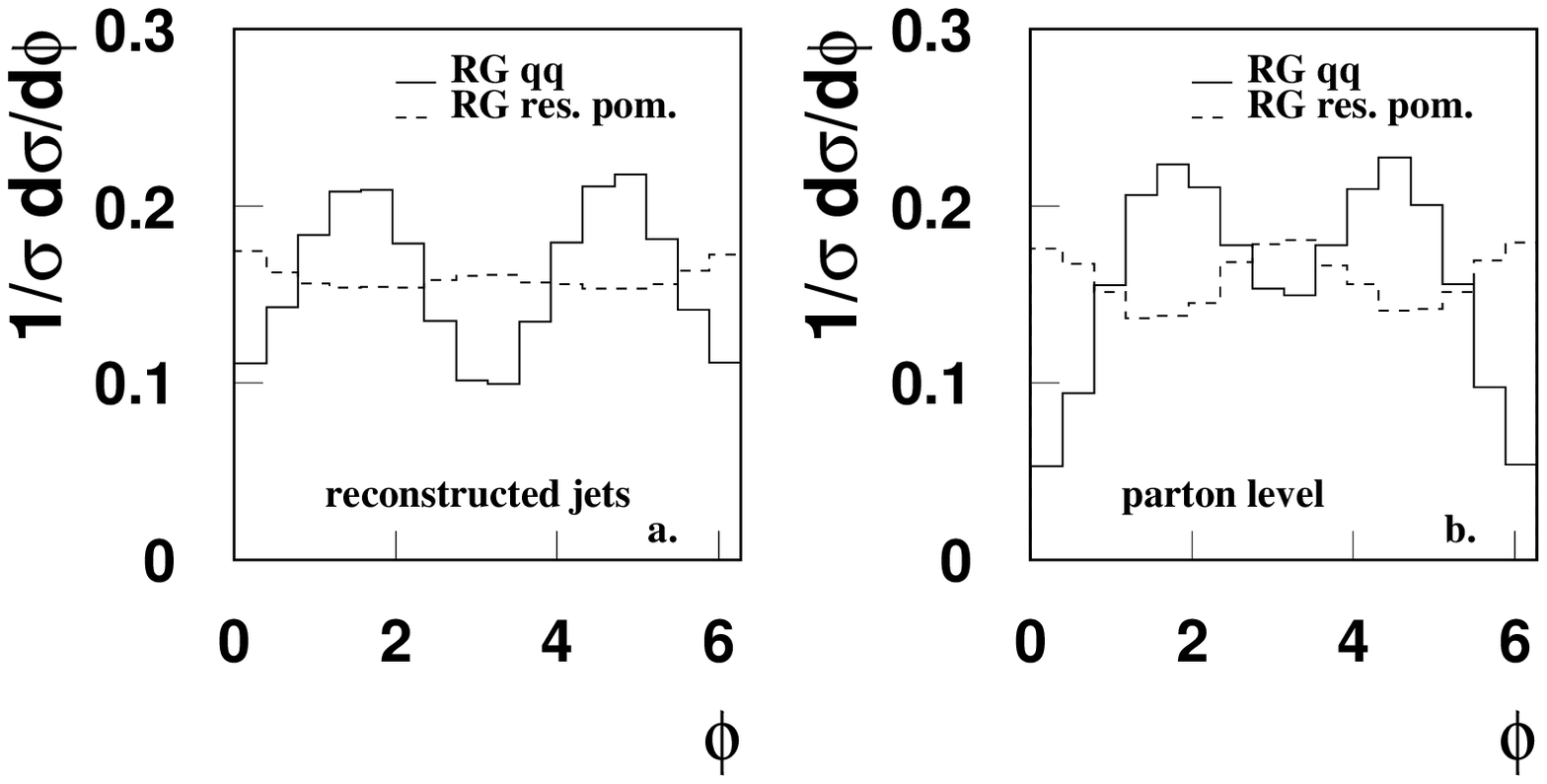,width=18cm,height=11cm}
\end{center}
\caption{$a.$ The $\phi$ asymmetry of one jet with the electron plane for 
high $p_T$ di jet events
in the region $0.1 < y < 0.7$, $5 < Q^2< 80$ GeV$^2$
$\xpom < 0.05$ and $p_T^{jet} > 2$ GeV. The solid line shows the 
prediction from the two gluon exchange mechanism
 after jet reconstruction at the hadron level.
The dashed line shows the $\phi$ dependence from a BGF type process in
diffraction (one gluon exchange). In $b.$ the $\phi$ asymmetry of the 
quark  with the electron plane is shown for comparison.
 The predictions are obtained with the RAPGAP Monte Carlo.
\label{dijet_phi}}
\end{figure}

The most striking feature of the perturbative QCD calculation of diffractive
$q\bar{q}$ final state is the $\phi$ asymmetry between the lepton and the quark
plane in the $\gamma^* p$ center of mass system. It is
difficult to identify the quark - jet at hadron level, 
therefore the jet with the largest $p_T$ is used.
The azimuthal asymmetry of the two gluon exchange mechanism 
obtained after jet reconstruction
 is  shown in Fig.~\ref{dijet_phi}, 
where also a comparison with the azimuthal asymmetry expected from a 
diffractive BGF process with one gluon exchange (from a resolved pomeron)
 is given. Even at the hadron level the 
difference between the two approaches is clearly visible.

\par
\par
However one has to be careful using this pQCD description of high $p_T$ dijet
production, since it is expected to be dominant only in a region where $Q^2 \gg
p_T^2$. In other regions of phase space where $Q^2 \sim p_T^2$ the contribution
from $q \bar{q} g$ final states are expected to become dominant. Such a
calculation is just being performed \cite{Bartels_qqg}.
\par
\begin{figure}[htb]
\begin{center}
\epsfig{figure=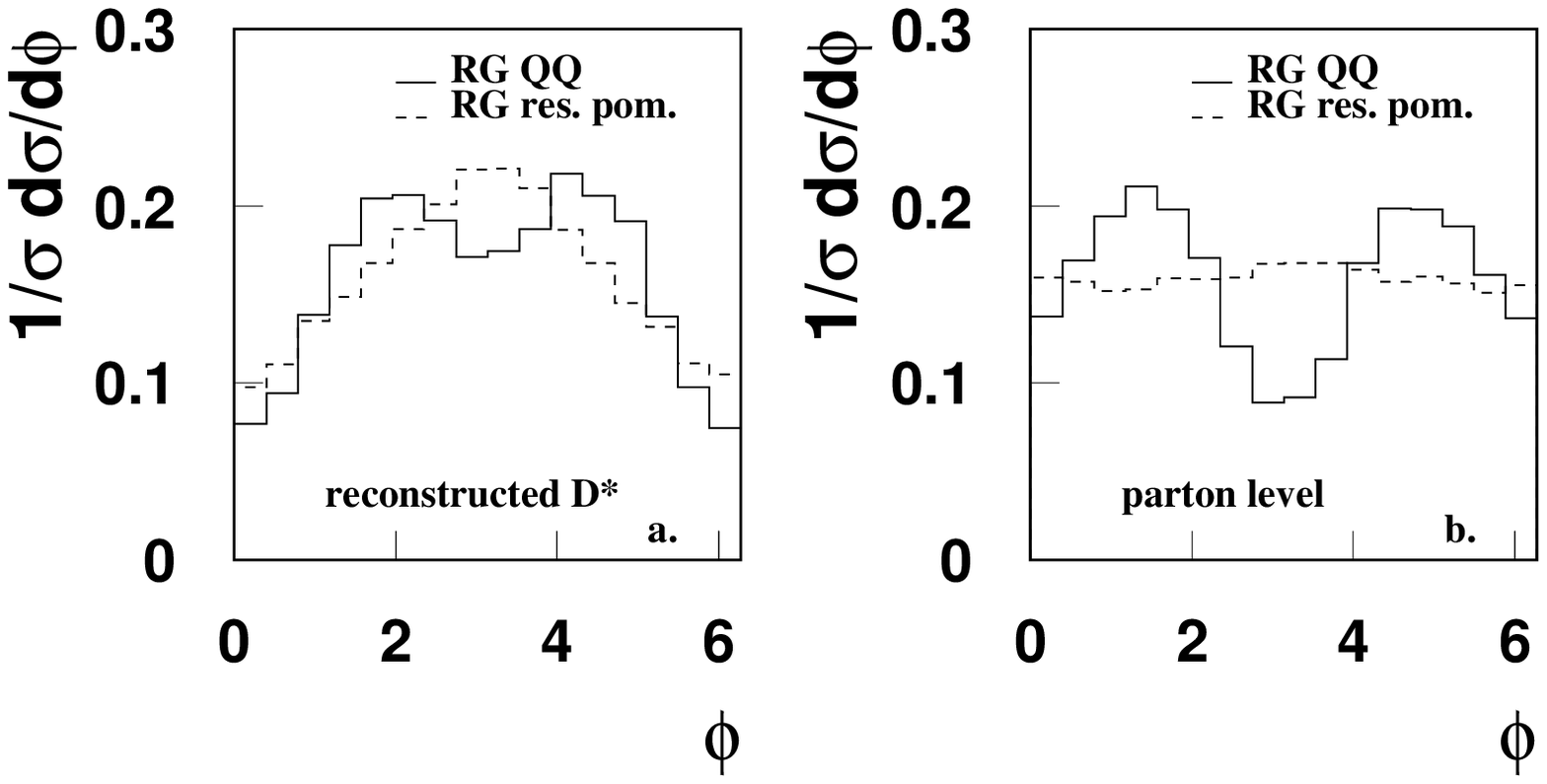,width=18cm,height=11cm}
\end{center}
\caption{$a$. The $\phi$ asymmetry of the $D^*$ jet with the electron plane in the 
kinematic region $0.06 < y < 0.6$, $2 < Q^2< 100$ GeV$^2$
$\xpom < 0.05$, $p_T^{D^*} > 1$ GeV and $|\eta^{D^*}| < 1.25$.
The solid line shows the 
prediction from the two gluon exchange mechanism after hadronization.
The dashed line shows the $\phi$ dependence from a BGF type process in
diffraction (one gluon exchange). In $b$ the $\phi$ asymmetry of the quark
  with the electron plane is shown.
 The predictions arc obtained with  the RAPGAP Monte Carlo.
\label{charm_phi}}
\end{figure}

As the calculation of a diffractive $q\bar{q}$ state can also be extended to
heavy quark production \cite{Lotter_charm,Diehl_charm},
 the difficulty of identifying
high $p_T$ di-jets may be avoided by the observation of $D^*$ mesons.  
In Fig.~\ref{charm_phi}$a$ the $\phi$ asymmetry is shown for $D^*$ 
mesons produced
by the two gluon exchange mechanism and compared to the prediction from a BGF
process in the resolved pomeron picture, where the  kinematic region is 
specified by: $0.06 < y < 0.6$, $2 < Q^2< 100$ GeV$^2$
$\xpom < 0.05$, $p_T^{D^*} > 1$ GeV and $|\eta^{D^*}| < 1.25$.
This process may thus also
be used to differentiate between the two approaches. One should note that the
different $\phi$ distribution observed here, as compared to ones from the jets,
is due to 
the cuts in the laboratory system used by the experiments to identify the $D^*$
meson. Without the $p_T^{D^*}$ cut, the $\phi$ distribution looks the same as
for the jets. In Fig.~\ref{charm_phi}$b$ the $\phi$ asymmetry at parton level
without the acceptance cuts is shown.

\subsection{Vector-meson production}

\begin{figure}[htb]
\begin{center}
\epsfig{figure=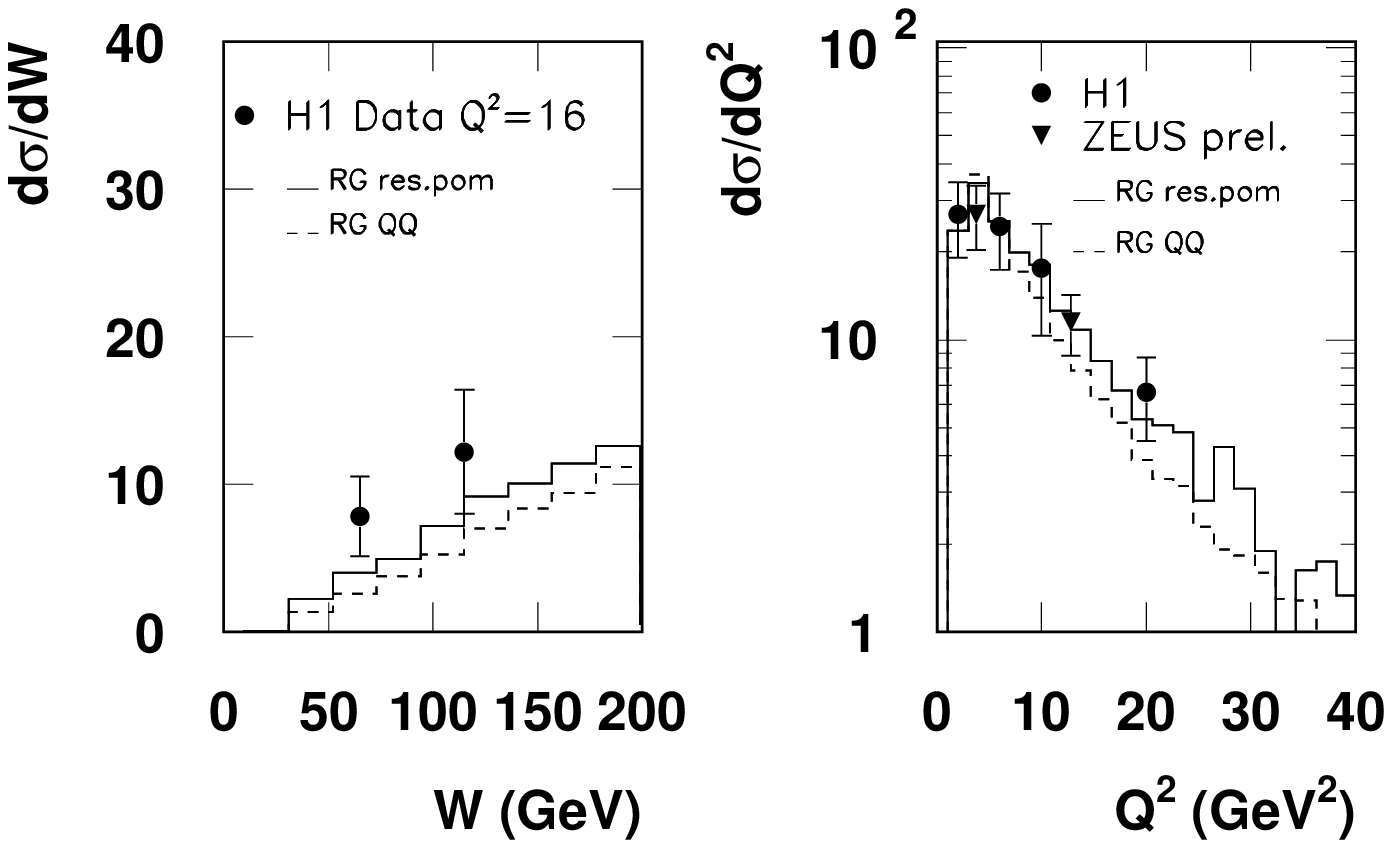,width=17cm,height=10cm}
\end{center}
\caption{The $\gamma* p$ cross section for exclusive $J/\psi$ production,
as a function of $W$ ($a.$) and as a function of $Q^2$ ($b.$). The solid
line is the RAPGAP prediction using the H1 $F_2^{D(3)}$ parameterization
\protect\cite{H1_F2D3_97} (fit 2).
The dashed line is the prediction from the pQCD calculation using the model of
$J/\psi$ production described in the text.
 In all predictions the charm mass was set
to $m_c = 1.5$ GeV.
The H1~\protect\cite{Jpsi_H1_dis} 
and ZEUS preliminary~\protect\cite{Jpsi_zeus_prelim} data are shown. 
In $b.$ the points at $Q^2< 10$ GeV$^2$  
are preliminary H1 data~\protect\cite{Jpsi_h1_prelim}. 
\label{jpsi}}
\end{figure}

The cross section for vector meson production depends 
crucially  on the underlying subprocess. For the calculation the charm mass
was set to $m_c=1.5$ GeV. 
First I shall describe $J/\psi$ production using a
 recent parameterization
of $F_2^{D(3)}(\xpom,\beta,Q^2)$ of the H1 collaboration \cite{H1_F2D3_97}.
This parameterization is based on
 a significant diffractive gluon density. Within
the model described before and implemented in RAPGAP,  
 $J/\psi$ production at large $Q^2$ as measured by 
H1~\cite{Jpsi_H1_dis} and ZEUS~\protect\cite{Jpsi_zeus_prelim}
can be well described, both as a function of $Q^2$ and  
$W$ (Fig.~\ref{jpsi}, solid line). It is remarkable that 
 a $\sim 1/Q^4$ dependence of the photon proton cross section
appears consistent with the data,
 which is usually interpreted as a higher twist effect.  Using
diffractive parton densities, this $Q^2$ dependence follows naturally from
the $\beta$ dependence of the structure function $F_2^{D(3)}$, since as 
shown before, changing $Q^2$ is equivalent with changing $\beta$ for 
a fixed mass of the vector-meson.
\par
In the perturbative QCD approach heavy quarks can be produced and, as mentioned
before, the cross section is essentially proportional to the gluon density
squared. For heavy quarks no restriction on the $p_T$ of the quarks is
necessary, and therefore the model of vector meson production described above 
can
be applied. In Fig.~\ref{jpsi} the prediction from this calculation using the
GRV HO parameterization of the gluon density in the proton is shown
with the dashed line. The cross
section agrees rather well with the data, both in shape and normalization.The
calculation agrees well with the data and also with the approach using
a parameterization of diffractive parton densities. However one should keep in
mind that the diffractive gluon density obtained from scaling violations of
$F_2^{D(3)}$ is only poorly constraint, so that there is still a significant
normalization uncertainty.

\par
A smaller value for the charm quark mass gives a larger phase space and a larger
cross section in addition to a change in the scale for $\alpha_s$ and 
for the parton distribution
functions which is essentially set by the charm quark mass.
Different choices of the charm mass result in an uncertainty in the overall
normalization,
 which has been estimated: 
$\sigma_{J/\psi}^{m_c=1.5} / \sigma_{J/\psi}^{m_c=1.35} \simeq 0.6$. A similar
uncertainty in normalization is also found for the color singlet model of
inelastic charm production, including the non relativistic wave function of the
$J/\psi$ - meson \cite{Kraemer_Jpsi}. 
\par
Given these uncertainties, both approaches are in fair agreement. If transverse
and longitudinally produced vector meson can be separated, differences
between the two approaches should show up, since in the perturbative calculation
the longitudinal part becomes large at large $Q^2$, which is not expected in the
resolved pomeron approach.
\par
One should note that the energy dependence of the cross section in the
perturbative approach emerges naturally from the gluon density in the proton,
whereas  using a parameterization of diffractive parton densities, the $x_{\PO}$
dependence has been inserted by hand to fit $F_2^{D(3)}$.
However it is a success, that the energy dependence obtained from $F_2^{D(3)}$
can be used to describe also $J/\psi$ production. 

\section{Summary}
The implementation of very different models for diffraction in deep inelastic
scattering has been described. It is found that available data 
on hadronic final state properties are reasonably
well described by the resolved pomeron model using a parameterization of
diffractive parton densities, by the soft color interaction model and the
perturbative QCD calculation involving two gluon exchange. It is shown that the
different models can be distinguished, if differential distributions are
considered. One particular example is diffractive charm production in deep
inelastic scattering, where the $\phi$ dependence of the cross section shows a
very different behavior in the two gluon exchange model compared to the
resolved pomeron model.
\par
It has been also shown that vector meson production can be nicely described
within a simple approach, both using the resolved pomeron model and the pQCD
calculation via two gluon exchange.
\par
Model predictions for the proton dissociative system have been presented in two
different approaches, within RAPGAP and the soft color interaction approach
implemented in LEPTO~6.5. Both models gave similar predictions.
\par
As more precise data on the hadronic final state in deep inelastic diffraction
are expected soon, it might be possible to distinguish and separate the various
approaches to diffraction and new and interesting insights into the structure of
the proton are expected.

\section*{Acknowledgements}
It was a great pleasure to participate at this interesting workshop. I am
grateful to the organizers A. Santoro and A. Brandt for this lively workshop and
the stimulating atmosphere. I am grateful to M. Erdmann, J. Gayler,
G. Ingelman and L. J\"onsson for 
careful reading of the manuscript.
\bibliographystyle{prsty} 
\bibliography{habilit}

%
%

%
%

\end{document}